\documentclass[aps,prb,amsfonts,amssymb,twocolumn,showkeys,showpacs,superscriptaddress,nofootinbib,10pt]{revtex4}

\usepackage{graphicx}
\usepackage{amsfonts}
\usepackage{amssymb}
\usepackage{amsmath}
\usepackage{verbatim}

\begin{document}

\title{The dynamically asymmetric SQUID: M\"unchhausen effect}

\author{A.U. Thomann}
\affiliation{Theoretische Physik, ETH Zurich, CH-8093 Zurich, Switzerland}
\author{V.B. Geshkenbein}
\affiliation{Theoretische Physik, ETH Zurich, CH-8093 Zurich, Switzerland}
\affiliation{L.D. Landau Institute for Theoretical Physics, 117940 Moscow, Russia}
\author{G. Blatter}
\affiliation{Theoretische Physik, ETH Zurich, CH-8093 Zurich, Switzerland}

\begin{abstract}
We report on a complex zero-temperature decay channel of a classical object in a metastable state coupled to a quantum degree of freedom. This setting can be realized in a dc-SQUID where both Josephson-junctions have identical critical currents $I_c$ but feature 
strongly asymmetric dynamical parameters; more precisely, selecting both parameters $C$ and $1/R$ adequately large for one and small for the other junction makes the first junction behave essentially classical but lets quantum effects be present for the second one. The decay process is initiated by the tunneling of the quantum junction, which distorts the trapping potential of the classical junction; the metastable state of the latter then becomes unstable if the distortion is large enough. We present the 
dynamical phase diagram of this system providing the dependence of this decay channel on the 
external bias current $I$ and on the coupling strength between the two junctions, determined by the loop inductance $L$.
\end{abstract}



\date{\today}

\maketitle

\section{Introduction}
\label{sec:introduction}

It is a basic feature of the classical world that a massive object residing in a metastable potential 
well cannot decay and, at zero temperature, is determined to reside in the well. There are, 
however, situations, where this doctrine is no longer adequate, namely if the system consists of two 
parts, a classical and a smaller quantum one. In that case, a decay process is possible under certain 
circumstances: The decay sequence is initiated by the tunneling of the quantum 
degree of freedom; through its coupling to the classical part the trapping potential is distorted, 
eventually turning the metastable state into an unstable one, provided that the distortion 
is large enough. We name this decay process the 'M\"unchhausen effect' after 
the famous baron telling the story of pulling himself out of a swamp by his own hair.

\begin{figure}[htb]
\centering
\includegraphics{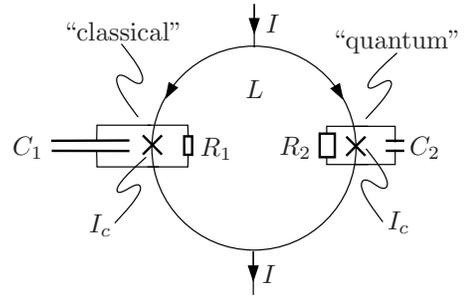}
\centering
\caption{Schematics of the dynamically asymmetric dc-SQUID. Two Josephson junctions 
with equal critical current $I_{c}$ but strongly asymmetric capacitances ($C_{1}\gg C_{2}$) and
resistances $R_{j}$ are integrated in a biased (current $I$) superconducting loop 
with symmetric inductance $L$.} 
\label{fig:setup}
\end{figure}

The above situation can be implemented experimentally in a dynamically 
asymmetric dc-SQUID; it is this specific realization we will study in detail in this paper, see 
Fig.\ \ref{fig:setup}. The dynamical degrees of freedom in the dc-SQUID are the 
gauge-invariant phase differences $\varphi_{j},\  j=1,2$, across the two Josephson junctions. 
The potential energy (of a single Josephson junction) is given by 
$\mathcal{V}_{j}=E_{J}(1-\cos\varphi_{j}),\ j=1,2$, 
involving the Josephson energy $E_{J}=\Phi_{0}I_{c}/2 \pi c$ (with the flux unit $\Phi_{0}=hc/2e$ and 
the critical current $I_{c}$ of the junction). The kinetic energy reads $\mathcal{T}_{j}=(\hbar/2e)^2
C_{j}\dot\varphi_{j}^2/2$, where the capacitances $C_{j}$ assume the role of effective masses. 
Hence, a SQUID featuring two Josephson junctions with equal critical current $I_{c}$ but adequately 
chosen and strongly asymmetric capacitances, one big and one small ($C_{1}\gg C_{2}$),
 effectively provides us with a classical and a quantum degree of freedom. 
The fabrication of large, classical junctions is easily achieved today; however, this is not the case for 
small quantum junctions whose realization is more difficult. Nevertheless, small junctions exhibiting 
quantum tunneling \cite{Martinis:1987,Balestro:2003} and quantum coherence \cite{Friedman:2000,Chiorescu:2003} 
can be routinely fabricated today.

The decay process of the biased dynamically asymmetric SQUID
proceeds in the following manner: The bias $I$, leading to a term $\propto -I (\varphi_{1}+\varphi_{2})$
in the potential, turns a stable state of the washboard potential into a metastable one. 
As junction 1 features a large capacitance, we assume its dynamics to be strictly 
classical. If the bias $I$ is large enough, a (imaginary time) 
decay process involving only the quantum junction (at constant $\varphi_{1}$) is enabled.
This phase slip leads to the entry of magnetic flux into the ring.
Given the inductive coupling $\propto1/L$  (inductance $L$),  
the current through the classical junction is enhanced and it may eventually become 
overcritical, thus decaying via a classical real time trajectory. 

In the following, we specify the setup (Sec.\ \ref{sec:setup}). In Sec.\ \ref{sec:pd},  
we describe the decay process in more detail and present the resulting dynamical phase diagram. 
We finish with a few concluding remarks in Sec.\ \ref{sec:conclusions}.

\section{Setup}
\label{sec:setup}

We start from the capacitively shunted junction model (CSJ), where the 
dc-SQUID, biased with a current $I$, is described by the Lagrangian
\begin{multline}
   \mathcal{L}=\sum_{j=1}^2 \bigg[\left(\frac{\Phi_{0}}{2\pi c}\right)^2\frac{C_{j}}{2}\dot\varphi_{j}^2 
   -E_{J}(1-\cos\varphi_{j})  \bigg]\\
   +\frac{\Phi_{0} I}{2 \pi c} \frac{\varphi_{1}+\varphi_{2}}{2}-
   \left(\frac{\Phi_{0}}{2\pi c}\right)^2\frac{(\varphi_{1}-\varphi_{2})^2}{2 L};
\end{multline}
here, we have assumed that the inductance $L$ of the SQUID is symmetrically distributed.
The Lagrangian $\mathcal{L}$ generates the equations of motion
\begin{equation}
   m_{j}\ddot\varphi_{j}+\eta_{j}\dot\varphi_{j}=-\partial_{\varphi_{j}}v(\varphi_{1},\varphi_{2}),
\label{eq:motion}
\end{equation}
with the `masses' $m_{j}=\Phi_{0}C_{j}/2\pi c I_{c}$ and where we have added the dissipative 
terms $\eta_{j}\dot\varphi_{j}$ with the damping parameters $\eta_{j}=\Phi_{0}/2 \pi c I_{c}R_{j}
\propto 1/R_{j}$, the normal resistances of the junctions;
the potential (illustrated in Fig.\ \ref{fig:v}) is given by
\begin{multline}
   v(\varphi_{1},\varphi_{2})=2-\cos\varphi_{1}-\cos\varphi_{2}\\
   -i (\varphi_{1}+\varphi_{2})+\frac{k}{2}(\varphi_{1}-\varphi_{2})^2,
\label{eq:v}
\end{multline}
with the dimensionless current $i=I/2 I_{c}$, the coupling constant $k=\Phi_{0}/2 \pi c I_{c}L$ 
and where energy is measured in units of $E_{J}$.

Residing in a symmetric ($\varphi_{1}=\varphi_{2}=\arcsin i$) metastable state of the 
potential $v(\varphi_{1},\varphi_{2})$ at finite bias current $i$, the classical version of 
the system described through Eq.\ \eqref{eq:motion} cannot decay at zero temperature. 
Here, we are investigating the case where junction one, featuring a large capacitance 
$C_{1}$, is assumed to behave strictly classical, whereas the dynamics of junction 2 is 
characterized by large quantum fluctuations.\footnote{Hence, a description via an equation 
of motion, Eq.\ \eqref{eq:motion}, is not suitable for the quantum junction and dissipative 
effects have to be included through a path integral formalism.} This can be achieved 
through a suitable choice of parameters, i.e., small capacitance $C_{2}$, while 
keeping $E_{J}\gtrsim E_{C2}=e^2/2 C_{2}$ such that we remain in a quasi-classical regime.

In the system under consideration, different scenarios can arise depending on the 
strength of the dissipation as quantified by the dimensionless damping parameter 
\begin{equation}
\alpha_{j}=(2 R_{j}C_{j}\omega_{pj})^{-1},\ j=1,2,
\end{equation}
where, at $i=0$, the plasma frequency $\hbar\omega_{pj}=(8 E_{J}E_{Cj})^{1/2}$. 
The simplest case is the overdamped situation $\alpha_{1},\alpha_{2}>1$,
where the dynamics of the classical junction is viscous and both relaxation and tunneling 
of the quantum junction are incoherent \cite{Leggett:1987, Caldeira:1981, Caldeira:1983}. We will analyze this situation 
in detail in the next section; the obtained results are also relevant for other choices of 
parameters, c.f. below.

\section{Decay process and phase diagram}
\label{sec:pd}

\begin{figure}[htb]
\centering
\includegraphics{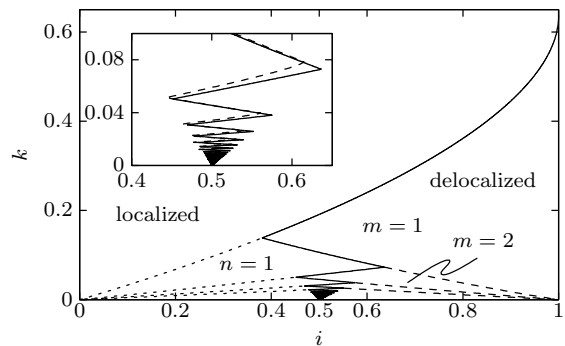}
\centering
\caption{Dynamical phase diagram of the dynamically asymmetric dc-SQUID. The critical 
line (solid) separates regions in the $i$-$k$-plane 
where the SQUID enters a continuous finite voltage 
state (delocalized) from regions where the phases remain localized. On the lower-bias side of the critical 
line, the dotted lines mark the entry of a additional flux into the SQUID loop. 
On the higher-bias side of the critical line, the dashed lines show how many
flux units are at least needed to render
the SQUID unstable. The inset shows 
a close up on the critical line and additionally displays the exact numerical solution (dashed).} 
\label{fig:pd_vv}
\end{figure}

In the following, we determine 
for which currents $i$ and coupling constants $k$ a zero temperature 
decay of a symmetric metastable state 
($\varphi_{1}=\varphi_{2}=\arcsin i$, up to an arbitrary multiple of $2\pi$) 
is allowed in the interferometer potential $v(\varphi_{1},\varphi_{2})$. 
The result is displayed in a dynamical phase diagram in the $i$-$k$-plane,
see Fig.\ \ref{fig:pd_vv}, 
where the critical line $i_{c}(k)$ separates regions where this decay is prohibited (localized state) 
from regions where it is allowed (delocalized).

Assuming junction 1 to behave strictly classical, 
(i.e. considering the limit of very large $C_{1}$), 
a quantum decay of the SQUID in a metastable state can only occur at fixed $\varphi_{1}$, i.e. 
through an imaginary-time trajectory of $\varphi_{2}$ in the effective potential $v_\mathrm{eff}
(\varphi_{2})=v(\varphi_{1}=\mathrm{const.},\varphi_{2}$). 
We will adopt this approximation for all tunneling processes throughout the discussion.

In the case $\alpha_{1},\alpha_{2}>0$ the preparation in the assumed symmetric 
initial state is straightforward:
A SQUID is cooled close to zero temperature at zero external magnetic field before a bias 
current is ramped to the desired value $i$. The ramp needs to be fast enough, 
such that the final current is reached before the
decay of the quantum junction; any initial potential energy is dissipated as the phases 
relax to the bottom of the metastable well. 

For undercritical currents $i<1$ the described initial state, $\varphi_{1}=\varphi_{2}=\arcsin i$, 
is stable against a decay involving the classical junction; however it is, for $k<i/(\pi-\arcsin i)$, 
unstable with respect to a quantum decay of $\varphi_{2}$ since the minimum of 
$v_\mathrm{eff}(\varphi_{2})$ near 
$\varphi_{2}\approx 2\pi$ is lowered below the initial one
and one or more phase slips of $\varphi_{2}$ are possible.
In order to determine whether the SQUID will remain in a stable or enter a finite voltage state 
at given $i,k$
we proceed in two steps: First, we have to determine where the successive 
quantum tunneling of $\varphi_{2}$ comes to a halt, 
i.e. which side minimum is quantum-stable. This is equivalent to finding the global minimum 
of the effective potential $v_\mathrm{eff}(\varphi_{2})$. Fixing 
the phase across the classical junction $\varphi_{1}=\arcsin i$,
 one immediately sees that 
the quantum-stable minimum is the one near $\varphi_{2}\approx 
2 \pi n$ if $n$ is the largest integer such that
\begin{equation}
   k<\frac{i}{(2 n-1)\pi-\arcsin i}.
   \label{eq:approx-globmin}
\end{equation}

\begin{figure}[htb]
\centering
\includegraphics[width=5.5cm]{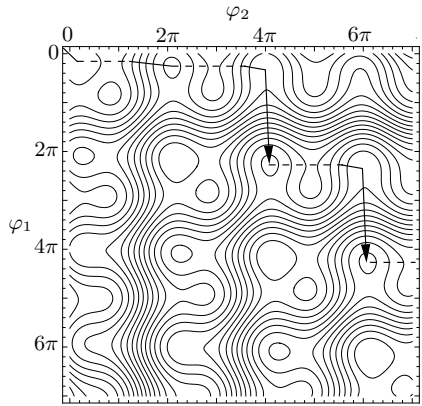}
\centering
\caption{Illustration of the decay sequence of the dynamically asymmetric SQUID ($i=0.5,\ 
k=0.04$). The initial metastable well is unstable w.r.t. the macroscopic quantum tunneling 
of the small junction 2. A continuous sequence of phase slips takes the system to a state 
which is classically unstable. In the following, the (classical) relaxation of the large junction 
1 and the quantum decay of junction 2 alternate and lead to a finite voltage state of the SQUID.} 
\label{fig:v}
\end{figure}

The sequence of phase slips of the quantum junction leads to an accumulation of magnetic 
flux in the SQUID loop, inducing a screening current. Consequently, the current through junction 
2 is reduced whilst that through junction 1 is increased. The magnitude of the induced current 
depends strongly on the coupling constant $k$; for given $n$, 
it is only large enough to drive the classical junction overcritical if 
\begin{equation}
  k > k_{c,n}(i)\approx\frac{1-i}{(2n-1/2)\pi+\arcsin(2i-1)}.
   \label{eq:approx-hesse}
\end{equation}
If this is the case, the SQUID enters a finite voltage state where quantum tunneling of junction 2 
(an approximate flux unit enters the loop)
and classical relaxation of junction 1 (flux leaves the loop) interchange sequentially (see 
Fig.\ \ref{fig:v}). If $k<k_{c,n}(i)$, the SQUID resides in a localized state and a further increase 
in $i$ is necessary to drive the system unstable.
The two conditions Eqs. \eqref{eq:approx-globmin} and \eqref{eq:approx-hesse} generate a web of 
crossing lines in the $i,k$-plane, determining the critical line $i_{c}(k)$ marking the dynamic transition 
from a localized to a delocalized state (Fig.\ \ref{fig:pd_vv}).
In the limit $k\rightarrow 0$, where the cosine in the potential Eq. \eqref{eq:v} becomes a small 
correction to the parabola, the critical line $i_{c}(k)$ 
approaches 1/2; this indicates that all current is redirected through junction 1 and delocalization 
takes place at $I=I_{c}$, the critical current of a single junction,

The simple arguments above have to be refined in order to obtain the precise location of 
the critical line $i_{c}(k)$. 
First, the condition of classical stability is 
  but the standard determination of the critical current of a dc-SQUID's 
asymmetric minimum \cite{Tsang:1975}. 
In our case, where classical stability along the $\varphi_{2}$-direction 
is guaranteed, the relevant set of equations is given by
\begin{eqnarray}
   \sin(\bar\varphi_{1}^{n})&=&i-k_{{c,n}}(\bar\varphi_{1}^{n}-
   \bar\varphi_{2}^{n}),\label{eq:hesse1}\\
   \sin(\bar\varphi_{2}^{n})&=&i+k_{{c,n}}(\bar\varphi_{1}^{n}-
   \bar\varphi_{2}^{n}),\label{eq:hesse2}\\
   \cos \bar\varphi_{1}^{n}\cos \bar\varphi_{2}^{n}
   &=& -k_{c,n}( \cos \bar\varphi_{1}^{n}+\cos \bar\varphi_{2}^{n}),
  \label{eq:hesse3}
\end{eqnarray}
and has to be solved (numerically) for $k_{c,n}(i)$ and  
$\bar\varphi_{1,2}^n$, the coordinates of the true minima 
near $\varphi_{1}=\arcsin i, \varphi_{2}=2 \pi n$.
Eq.\ \eqref{eq:approx-hesse} is
an approximate solution to Eqs.\ \eqref{eq:hesse1}-\eqref{eq:hesse3} in the limit of 
$k\ll 1$.
Second, as tunneling of the quantum junction might be enabled only after the 
relaxation of the classical junction to a minimum, condition
Eq.\ \eqref{eq:approx-globmin} has to be corrected to 
\begin{equation}
   k_{c,n}(i)=\frac{i}{(2 n-1)\pi-\bar\varphi_{1}^{n-1}},
   \label{eq:globmin}
\end{equation}
taking into account the change of the effective potential $v_\mathrm{eff}(\varphi_{2})$ upon 
a change in $\varphi_{1}$.
The exact numerical solutions 
of Eqs.\ \eqref{eq:hesse1}-\eqref{eq:hesse3} and \eqref{eq:globmin} are shown in the inset 
of Fig.\ \ref{fig:pd_vv}, where the approximate solution is seen to be rather precise.

A few remarks on the above results are in place.
First, the setup as described above may not be suitable for experimental investigations 
as an overdamped quantum junction suffers from strongly suppressed tunneling rates. However, 
the analogous results can be obtained by using two underdamped junctions and ramping the current sufficiently slowly, 
such that the junctions have dissipated all 
energy before crossing any line (solid or dotted) in the phase diagram Fig.\ \ref{fig:pd_vv}. 
The hallmark 
of the ``M\"unchhausen-effect'' in this case is the two different types of decay that 
finally initiate a 
running state: If the critical line is crossed on a part with positive slope, the quantum stable minimum 
is classically unstable immediately 
and a finite voltage state is initiated by the last quantum 
tunneling process. On the other hand, if the critical line is crossed on a part with negative slope, the 
quantum-stable minimum is still classically stable after the tunneling and needs to be turned 
overcritical by increasing the 
bias. Then, the finite voltage state is initiated by a purely classical decay. The difference between the 
two types of decay 
should be visible upon analyzing the decay histograms of multiple measurements, being broad 
for a quantum decay but narrow for a classical process.

Second, one has to take into account that for realistic systems, where the capacitance asymmetry 
is introduced by a capacitive shunt \cite{Steffen:2006}, the maximum asymmetry is limited. 
Corrections due to the finite mass of the classical junction might appear in the form of 
two-dimensional quantum tunneling, where junction 1 takes part in the tunneling process 
\cite{Morais-Smith:1994}. 

\section{Conclusions}
\label{sec:conclusions}

We have shown that 
a dc-SQUID can decay out of a symmetric metastable state, 
even if one of the junctions behaves \emph{fully} classical, provided the second junction shows 
quantum behavior. The ``M\"unchhausen-decay''
involves tunneling of the quantum junction, where magnetic flux accumulates in the 
superconducting ring and eventually redirects enough current through the classical junction 
as to drive it overcritical.
We thank A. Ustinov and A. Wallraff for fruitful discussions.

\end{document}